\documentclass[11pt]
{revtex4}

\usepackage{amscd}
\usepackage{amsmath}

\begin{document}

\title{ON THE WAVE FUNCTIONS AND ENERGY SPECTRUM  \\ FOR A SPIN 1
PARTICLE IN EXTERNAL COULOMB FIELD}
\author{V.V. Kisel  }
\author{E.M. Ovsiyuk}
\author{V.M. Red'kov }

\email{ redkov@dragon.bas-net.by, e.ovsiyuk@mail.ru}

\affiliation{Belarusian  State Pedagogical University \\
Mozyr State Pedagogical University \\
Institute of Physics, NAS
of Belarus}


\begin{abstract}

Quantum-mechanical system -- spin 1 particle in external Cou\-lomb
field is studied on the base of the matrix Duffin -- Kemmer --
Petiau formalism with the  use of the tetrad technique. Separation
of the variables is performed with the help of Wigner functions
  $D^{j}_{-m\sigma}(\phi, \theta,0); \;
\sigma = -1,0, +1$; $j$ and $m$ stand for quantum numbers
determining the square and
 the third projection of the total
angular momentum of the vector particle. With the help of parity
operator, the radial 10-equation system is  divided into two
subsystem of 4 and 6 equations that correspond to  parity $P=
(-1)^{j+1}$ and  $P= (-1)^{j}$ respectively. The system of 4
equation is reduced to a second order differential equation which
coincides with that arising in the case of a scalar particle in
Coulomb potential. It is shown that the 6-equation system reduces
to two different second order  differential equations for a "main"
\hspace{2mm} function. One main equation reduces  to a confluent
Heun equation  and provides us  with energy spectrum. Another main
equation is a more complex one, and any solutions for it are not
constructed.

\end{abstract}



\maketitle

\section{Introduction and separation of the  variables}

Many years ago, a very peculiar behavior of a spin 1 particle in presence of the external Coulomb field was noticed by I.E. Tamm
\cite{Tamm}. As far as we known the whole situation with  this system stays much the same.
In the present paper, we start examining the problem on the base of the
matrix Duffin -- Kemmer -- Petiau formalism
\cite{Petiau-1936}, \cite{Duffin-1938}, \cite{Kemmer-1939, Kemmer-1943}  with the use of the
tetrad generally covariant technique (for more details and references see in
\cite{Bogush-Otchik-Red'kov-1986, Red'kov-1998(3), Red'kov-1998(4), Book(1)}), it turns  out to be more convenient than a common Proca  tensor approach
\cite{Proca-1938, Proca-1946}). Choosing  a  diagonal
spherical tetrad according to
\begin{eqnarray}
 dS^{2} = \  dt^{2} -  dr^{2}  -
r^{2} (d\theta^{2} +  \sin ^{2}\theta d\phi ^{2}) \; , \nonumber
\\
e^{\alpha}_{(0)}=(1 , 0, 0, 0) \; , \qquad e^{\alpha}_{(3)}=(0, 1,
0, 0) \; , \nonumber
\\
e^{\alpha}_{(1)}=(0, 0, \frac {1}{r}, 0) \; , \;\;\;
e^{\alpha}_{(2)}=(1, 0, 0, \frac{1}{ r  \sin \theta})  \; ,
\label{1.1}
\end{eqnarray}

\noindent we   reduce the main D-K-P  stationary equation to  the form
\begin{eqnarray}
\left [ \;\beta ^{0} ( \epsilon  + {\alpha \over r} ) +i \; [
\beta ^{3}
\partial _{r}   +  {1 \over r}  ( \beta ^{1} j^{31}  +
\beta ^{2} j^{32} ) ]  +  {1  \over r} \Sigma _{\theta,\phi }  - M
   \right  ]  \Phi (x)  =  0 \; ,
\label{1.2}
\end{eqnarray}

\noindent where $\epsilon = E / (c \hbar),  \alpha = e^{2}/ (c \hbar ), M = mc / \hbar$
and $\Sigma _{\theta ,\phi }$ stands for  an angular
operator (its form means that we have here a generalized Schr\"{o}dinger -- Pauli  basis
\cite{Schrodinger-1938, Pauli-1939})
\begin{eqnarray}
\Sigma _{\theta ,\phi } =  \; i\; \beta ^{1}
\partial _{\theta } \;+\; \beta ^{2}\; {i\partial \;+\;i\;j^{12}\cos
\theta  \over \sin \theta  } \; . \nonumber
\end{eqnarray}

\noindent Spherical  waves with $(j,m)$ quantum numbers
should be constructed within the following general
 substitution (we adhere notation developed in
 Red'kov [16-22, 30];
before similar  techniques  was applied
by Dray \cite{Dray-1985, Dray-1986},
Krolikowski and Turski
\cite{Krolikowski-Turski-1986},
Turski \cite{Turski-1986}
;  many years ago such a tetrad basis was  used by
 Schr\"{o}dinger \cite{Schrodinger-1938}
 and Pauli \cite{Pauli-1939} when looking at the problem
of single-valuedness of wave  functions in quantum theory -- then the case of spin  $S=1/2$ particle was
 specified)
\begin{eqnarray}
\Psi (x) = \{  \;  \Phi_{0} (x) , \; \vec{\Phi}(x) , \; \vec{E}(x)
, \vec{H}(x) \; \} \; , \nonumber
\\
\Phi_{0}(x) =    e^{-iE t/ \hbar}\; \Phi_{0} (r) \; D_{0} \; ,
\qquad \vec{\Phi}(x)   = e^{-iE t/ \hbar}\; \left |
\begin{array}{l}
\Phi_{1} (r)\; D_{-1}  \\
\Phi_{2}(r) \; D_{0}   \\
\Phi_{3}(r) \; D_{+1}  \end{array} \right | \; , \nonumber
\\
\vec{E} (x)  = e^{-iE t/ \hbar} \; \left | \begin{array}{l}
E_{1}(r) \;  D_{-1} \\
E_{2}(r) \;  D_{0}   \\
E_{3}(r) \;  D_{+1}     \end{array} \right |  , \;\; \vec{H} (x) =
e^{-iE t/ \hbar} \; \left | \begin{array}{l}
H_{1}(r)  \; D_{-1}  \\
H_{2}(r)  \; D_{0}   \\
H_{3}(r)  \; D_{+1}    \end{array} \right |\; ; \label{1.3}
\end{eqnarray}

\noindent    short notation for  Wigner functions \cite{Wigner-1927} is used: $
D_{\sigma}= D_{-m,\sigma}^{j}(\phi,\;\theta,\;0) \; , \;\;
\sigma=0,\;+1,\;-1$. In accordance with  Pauli approach \cite{Pauli-1939} the quantum number
 $j$ takes values 0,1,2, ...
 With the help of recurrent formulas  \cite{Varshalovich-Moskalev-Hersonskiy-1975} (where $
\nu =  \sqrt{j(j+1)} \; , \;  a = \sqrt{(j-1)(j+2)} $)
\begin{eqnarray}
\partial_{\theta} \;  D_{-1} =   {1 \over 2} \; ( \;
a  \; D_{-2} -    \nu  \; D_{0} \; ) \; , \qquad
 \frac {m -
\cos{\theta}}{\sin {\theta}} \; D_{-1} = {1 \over 2} \; ( \; a \;
D_{-2} + \nu \; D_{0} \; ) \; , \nonumber
\\
\partial_{\theta}  \; D_{0} = {1 \over 2} \; ( \;
\nu  \; D_{-1} - \nu  \; D_{+1} \; ) \; , \qquad
 \frac {m}{\sin
{\theta}} \; D_{0} = {1 \over 2} \; (  \;  \nu \; D_{-1} +  \nu \;
D_{+1} \; ) \; , \nonumber
\\
\partial_{\theta} \; D_{+1} =
{1 \over 2} \; ( \; \nu  \; D_{0} - a  \; D_{+2} \; ) \; , \qquad
\frac {m + \cos{\theta}}{\sin {\theta}}  \; D_{+1} ={1 \over 2} \;
( \;
 \nu \; D_{0} + \ a  \; D_{+2} \; ) \;   ,
\label{1.4}
\end{eqnarray}

\noindent after simple algebraic calculation we arrive at
the  radial equations (for clarity  corresponding Proca tensor equations are written down as well)

\vspace{2mm}
$
 D^{b} \; \Phi_{ab} = M \; \Phi_{a}  \;
, $
\begin{eqnarray}
 - \;\;({d \over dr}  + \frac {2}{r}) \; E_{2} -
\frac{\nu}{r}\;  (E_{1} + E_{3}) = M \Phi_{0}  \; , \nonumber
\\
+ i (\epsilon + {\alpha \over r} ) \; E_{1} \;\; + i \;({d \over
dr} + \frac {1}{r}) \; H_{1} + i \frac{\nu}{r} \; H_{2} =  M
\Phi_{1} \; , \nonumber
\\
+ i ( \epsilon + {\alpha  \over r}) \; E_{2} \;\;  -
i\frac{\nu}{r}\; (H_{1} - H_{3}) =  M \Phi_{2}  \; , \nonumber
\\
+i(\epsilon + {\alpha  \over r}) \; E_{3} \;\; - i \;({d\over dr} +
\frac {1}{r})\; H_{3} - i \frac{\nu}{r} \; H_{2} =  M \; \Phi_{3}
\;   ; \label{1.5}
\end{eqnarray}

$ \ D_{a} \Phi_{b} - D_{b} \Phi_{a} =  M \; \Phi_{ab}
\; , $
\begin{eqnarray}
-i (\epsilon + {\alpha \over r}) \; \Phi_{1}\;\;  + \frac{\nu}{r}
\; \Phi_{0} - M E_{1} = 0 \;     , \; \nonumber
\\
-i (\epsilon +{\alpha \over r})  \; \Phi_{2} \;\; - \;{d \over  dr}
\; \Phi_{0} - M E_{2} = 0 \;    , \nonumber
\\
-i ( \epsilon  + {\alpha \over r} ) \; \Phi_{3} \;\;  +
\frac{\nu}{r} \; \Phi_{0}  \; - \; M E_{3} = 0 \;   , \; \nonumber
\\
 - i\; ({d \over dr} + \frac {1}{r}) \; \Phi_{1} -
i\frac{\nu}{r} \; \Phi_{2} \;  - \; M H_{1} = 0 \;  , \nonumber
\\
+i\frac{\nu}{r}\; (\Phi_{1} - \Phi_{3}) - M H_{2} = 0 \; ,
\nonumber
\\
  +i\;({d \over dr} + \frac {1}{r})\; \Phi_{3} +
i\frac{\nu}{r} \Phi_{2} - M H_{3} = 0 \; , \label{1.6}
\end{eqnarray}

\noindent where  $\nu = \sqrt{j(j+1)/2}$ (note the factor $1/\sqrt{2}$).

Concurrently with  $\vec{J}^{\;2}, J_{3}$ let us  diagonalize an
operator of spatial inversion
 $\hat{\Pi}$. After transition to spherical tetrad basis, and also to cyclic representation for D-K-P matrices
  $\beta^{a}$, for this discrete operator we get
  \begin{eqnarray}
\hat{\Pi}   =  \left | \begin{array}{cccc}
1  &  0  &  0  &  0    \\
0  &  \Pi_{3}   &  0  &  0   \\
0  &  0  &  \Pi_{3}  &  0     \\
0  &  0  &  0 &   -\Pi_{3}
\end{array} \right |      \hat{P} \; , \qquad
\Pi_{3}  =    \left | \begin{array}{rrr}
0  &  0  &  -1  \\
0  &  -1 &   0  \\
-1  &  0  &  0   \end{array} \right | \; . \label{1.7}
\end{eqnarray}

\noindent
Eigen-value equation  $\hat{\Pi} \Psi = P \; \Psi $  results in two different in parity states
\begin{eqnarray}
\underline{P=(-1)^{j+1}} , \qquad \Phi_{0} = 0 \; , \;\; \Phi_{3}
= - \Phi_{1} \; , \; \Phi_{2} = 0 \; , \nonumber
\\
E_{3} = - E_{1} \; , \; E_{2} = 0 , \; H_{3}  = H_{1} \; ;
\label{1.8}
\end{eqnarray}
\begin{eqnarray}
\underline{P=(-1)^{j}} \; , \;\; \qquad   \Phi_{3} =
\Phi_{1} \; , \;
E_{3} = + E_{1} \; , \; H_{3} = - H_{1} \; , \; \; H_{2} = 0 \; .
\label{1.9}
\end{eqnarray}

\noindent Correspondingly,
eqs.   (\ref{1.5}) -- (\ref{1.6}) give  4 and 6 equations

\vspace{3mm} $\underline{P=(-1)^{j+1}} $,
\begin{eqnarray}
+i ( \epsilon  + {\alpha \over r } ) \; E_{1} +  i ({d \over dr} +{
1 \over r}) H_{1} +
 i {\nu \over r} H_{2}
= M \Phi _{1} \; , \nonumber
\\
-i  ( \epsilon  + {\alpha \over r}) \; \Phi_{1}  = M E_{1} \; ,
\;\; -i ({d \over dr} + {1 \over r}) \Phi_{1}   = M H_{1} \; ,
\;\; 2i { \nu \over r }  \Phi_{1} = M H_{2}  \; , \label{1.10}
\end{eqnarray}

\noindent excluding  $E_{1}, H_{1}, H_{2}$  we get a second  order differential equation
for  $\Phi_{1}$
\begin{eqnarray}
\left [ \;{ d^{2} \over dr^{2} } + {2 \over r} {d \over d r}  +
(\epsilon + {\alpha \over r})^{2} -M^{2} - {j(j+1) \over r^{2}} \;
\right ]  \Phi_{1} = 0 \; , \label{1.11}
\end{eqnarray}

\noindent which  coincides with that
arising in the case of a scalar particle in Coulomb potential (it is the known fact that was noted  in Tamm's first paper
\cite{Tamm}. Its solution is well known and
provides us with the  following energy spectrum (in usual units)
\begin{eqnarray}
E  = { mc^{2}  \over  \sqrt{1 +  \alpha^{2} /  N^{2}  }}
 \;, \qquad
N = n + {1\over 2} + \sqrt{(j+1/2)^{2} - \alpha^{2}}
 \; .
\label{1.11'}
\end{eqnarray}

For states with parity  $P=(-1)^{j}$ we have the system

\vspace{3mm} $\underline{P=(-1)^{j}} $,
\begin{eqnarray}
({d \over dr} +{2\over r}) E_{2} + 2 {\nu \over r} E_{1} + M
\Phi_{0} = 0 \; , \nonumber
\\
\;\; + i ( \epsilon + {\alpha  \over r}) \;  E_{1} + i ( {d \over
dr} + {1 \over r}) H_{1}  - M \Phi_{1} = 0 \; , \nonumber
\\
+ i ( \epsilon + {\alpha \over r})  E_{2} -2i {\nu \over r}  H_{1}
- M \Phi_{2} = 0\; , \nonumber
\\
\;\; -i ( \epsilon + {\alpha \over r}) \; \Phi_{1} + {\nu \over r}
\Phi_{0} - M E_{1} =0 \; , \; \nonumber
\\
i ( \epsilon + {\alpha \over r}) \Phi_{2}  +  {d \over dr} \Phi_{0}
+ M E_{2} = 0\; , \;\; \;\; \nonumber
\\
i ({d \over dr}  +{1\over r}) \Phi_{1} + i {\nu \over r} \Phi_{2}
+ M H_{1}  = 0\; . \label{1.12}
\end{eqnarray}

\section{ The case of minimal value  $j=0$}

The states with minimal value  $j=0$ can be treated straightforwardly (also see in
\cite{Kisel-Tokarevskaya-Red'kov-2003}).
In this case, we should start with special substitution of the wave function
\begin{eqnarray}
\Phi_{0}(x) =  e^{-iE t/ \hbar}  \Phi_{0} (r) \; , \;
C(x)  = e^{-iE t/ \hbar}  C (r) \;   , \; C_{0}(x)
=e^{-i\epsilon t} C_{0}\; (r) , \nonumber
\\
\vec{C} (x) = e^{-iE t/ \hbar} \left | \begin{array}{c}
0  \\
 C_{2}(r)     \\
0  \end{array} \right |  , \qquad \vec{\Phi}(x)   = e^{-i\epsilon
t} \left | \begin{array}{c}
0   \\
\Phi_{2}(r)    \\
0  \end{array} \right | , \nonumber
\\
\vec{E} (x)  = e^{-iE t/ \hbar} \left | \begin{array}{c}
0  \\
E_{2}(r)   \\
0  \end{array} \right |  , \qquad \vec{H} (x) =  e^{}
\left | \begin{array}{c}
0  \\
H_{2}(r)  \\
0    \end{array} \right |\; . \label{2.1}
\end{eqnarray}

\noindent  The $\Sigma_{\theta, \phi}$   acts on this function as a zero operator, and parity
$P=(-1)^{0+1}= -1 $.  Now  the radial system is (for eliminating imaginary  unit $i$,
 we use slightly different variables:
 $\Phi_{0} = \varphi_{0}, \; -i  \Phi_{1} =
\varphi_{1}, \; -i \Phi_{2} = \varphi_{2} $)
\begin{eqnarray}
H_{2} = 0\; , \qquad  - ({d \over d r} +{2 \over r}) E_{2} = M
\varphi_{0}\; , \nonumber
\\
(\epsilon  + {\alpha \over r})  E_{2}   = M \varphi_{2} \; , \qquad
(\epsilon + {\alpha \over r}) \varphi_{2}  - {d \over d r}
\varphi_{0} = M E_{2} \; . \label{2.2}
\end{eqnarray}

\noindent From whence  it follows a second order  equation for  $E_{2}$
\begin{eqnarray}
\left [  { d^{2} \over dr^{2}} + {2 \over r} {d \over d r} - {2
\over r^{2}}+ ( \epsilon + {\alpha \over r})^{2} - M^{2} \right ]
E_{2} = 0 \; . \label{2.3}
\end{eqnarray}

\noindent With substitution  $E_{2}(r) = r^{-1} f(r)$, we get
\begin{eqnarray}
{d^{2} \over dr^{2}} f + ( \epsilon^{2} - M^{2} + {2 \alpha
\epsilon \over r} - {2 - \alpha^{2} \over r^{2}} ) f = 0 \; .
\label{2.3}
\end{eqnarray}

\noindent In  dimensionless variables
\begin{eqnarray}
x =  r \epsilon = {r E \over c \hbar } \; ,  \;\;
{M^{2} \over  \epsilon^{2}} = { m^{2} c^{4} \over E^{2}} = \lambda^{2} \; ,
\nonumber
\label{2.5}
\end{eqnarray}

\noindent it reads
\begin{eqnarray}
{d^{2}  \over dx^{2} } \; f\;  + \; ( \; 1 \; - \; \lambda ^{2} \;
+ \; { 2 \alpha  \over x} \; - \; { 2 -  \alpha^{2} \over x^{2} }
\; ) \; f  = 0 \; . \label{2.6}
\end{eqnarray}

\noindent With the substitution  $ f(x) = x^{a} e^{-bx} F (x)$,
for $F$  we obtain
\begin{eqnarray}
 x   F''  +   ( 2a - 2bx)   F'   +
 [ \;
{a(a-1) + \alpha^{2} - 2 \over x} +
(b^{2} +1 - \lambda^{2} ) x +
(2\alpha - 2ab ) \; ] F = 0 \; .
\nonumber
\label{2.7}
\end{eqnarray}

\noindent Requiring
\begin{eqnarray}
a = {1 \pm \sqrt{9 - 4 \alpha^{2}} \over 2} \; , \;\; b = \pm
\sqrt{\lambda^{2} - 1} = \pm { \sqrt{m^{2} c^{4} - E^{2}}
\over E } \; , \label{2.7}
\end{eqnarray}

\noindent the choice of upper signs in the formulas provides us with
good parameters for bound states, we get
\begin{eqnarray}
 x \;  F'' \; +  \; 2 ( a - b x) \;  F' \; + \;
2(\alpha - a b ) \;  F = 0 \; . \label{2.8}
\end{eqnarray}

\noindent
This equation is solved in confluent  hypergeometric functions. Let us specify this solution in detail.
With the use of expansion for  $F(x)$
\begin{eqnarray}
F (x) = \sum_{k=0}^{\infty}\; C_{k} \; x^{k} \; ,\qquad
F' = \sum_{k=1}^{\infty} \; k C_{k}\;  x^{k-1} \; , \;\; F'' =
\sum_{k=2}^{\infty} \; k(k-1) C_{k} \; x^{k-2} \; , \nonumber
\end{eqnarray}

\noindent we get
\begin{eqnarray}
\sum_{k=2}^{\infty} \; k (k-1) C_{k}\; x^{k-1} \; + \; 2a \;
\sum_{k=1}^{\infty} \; k  C_{k} \; x^{k-1}  -
 2b  \;
\sum_{k=1}^{\infty} \; k  C_{k} \; x^{k} \; + \; 2(\alpha - a b)
\;  \sum_{k=0}^{\infty} \; c_{k} \; x^{k} = 0 \; , \nonumber
\end{eqnarray}

\noindent from whence it follows
\begin{eqnarray} \;
[ \; 2aC_{1} + 2 (\alpha - a b) \; ] x^{0} \; + \; [ \; 2 C_{2} +
2a \; 2C_{2} - 2 b \; C_{1} + 2(\alpha -a b) C_{1} \; ] \; x +
\nonumber
\\
+ \; \sum_{n=2}^{\infty} \; [ \; n(n+1) C_{n+1} + 2a (n+1) C_{n+1}
- 2bnC_{n} + 2 (\alpha -a b ) C_{n} \; ] \; x^{n} = 0 \; .
\nonumber
\end{eqnarray}

\noindent Therefore, we arrive at recurrent formulas
\begin{eqnarray}
C_{1} = - (\alpha -a b) \; C_{0} = 0 \; , \nonumber
\\
 C_{2} \;2 \;  (1  + 2 a )  = 2 \; [ b - (\alpha -a b)] \; C_{1} = 0
, \;  n = 2,3,4, ... , \nonumber
\\
C_{n+1} \; (n+1)\;  (  n + 2a )  =  2 \; [ n\; b\;  -\;   (\alpha
-a b )\; ]\; C_{n}   = 0 \; . \label{2.9}
\end{eqnarray}

\noindent To get the polynomial we must require
\begin{eqnarray}
C_{N+1} = 0 \; \Longrightarrow \; \; [\; N \; b\;  - \;   (\alpha
- a b )\; ] \; =0 \; . \nonumber
\end{eqnarray}

\noindent This quantization rule
gives
\begin{eqnarray}
{ \alpha - a b \over b } = N \; ,\qquad
a = {1 + \sqrt{9 - 4 \alpha^{2}} \over 2} \; , \;\; b =   { \sqrt{m^{2}c^{4}  - E^{2}} \over
E } \; ,
\nonumber
\end{eqnarray}

\noindent so that
\begin{eqnarray}
{ 2 \alpha \epsilon - (1 + \sqrt{9 -4\alpha^{2}}) \sqrt{m^{2}c^{4}
- E^{2} }\over 2 \sqrt{m^{2}c^{4} - E^{2} }} = N \; ;
\nonumber
\label{2.11}
\end{eqnarray}

\noindent its solution is
\begin{eqnarray}
E = {mc^{2} \over  \sqrt{1 + \alpha^{2}  / (\Gamma + N )^{2}   } } \;,
\qquad 2\Gamma = 1 + \sqrt{9 -4\alpha^{2}}
 . \label{2.12}
\end{eqnarray}

\section{  Lorentz condition in presence  of Coulomb
potential}

As known for a  massive spin 1 particle there must exist
a generalized Lorentz condition. Let us specify it for the problem under consideration.
In the   Proca  tensor  equations for  the vector particle
 \begin{eqnarray}
D_{\alpha } \; \Psi _{\beta } - D_{\beta } \; \Psi _{\alpha } = M
\; \Psi _{\alpha \beta }\; ,\qquad D^{\alpha } \;\Psi _{\alpha
\beta } = M\; \Psi _{\beta } \; , \label{3.1}
\end{eqnarray}

\noindent where  $D_{\alpha }  = \nabla _{\alpha } + i\; (e/c)\;
A_{\alpha }$, let us act by the operator  $D_{\alpha
}$ on the second equation in (\ref{3.1}), we get
\begin{eqnarray}
 ( \nabla _{\alpha } \; + \; i \;{ e\over c}   A_{\alpha }) \; \Psi
^{\alpha } =  {ie \over 2c  M }
 \; F_{\alpha \beta } \; \Psi ^{\alpha \beta }  \; .
\label{3.2'}
\end{eqnarray}

\noindent  This Lorentz condition should be translated to tetrad form.
To this  end, instead of   $\Psi ^{\alpha }$
and   $\Psi ^{\alpha \beta}$  one should introduced their tetrad components
\begin{eqnarray}
\Psi ^{\alpha } = e^{(a)\alpha } \; \Psi _{(a)} \; , \;\;\; \Psi
^{\alpha \beta }  = e^{(a)\alpha } \; e^{(b)\beta } \; \Psi
_{(a)(b)}\;. \nonumber
\end{eqnarray}

\noindent Correspondingly, eq.  (\ref{3.2'}) takes the form
\begin{eqnarray}
( e^{(a)\alpha }_{;\alpha } \; \Psi _{a} \; +  \; e^{(a)\alpha }
\; \partial _{\alpha } \; ) \; \Psi _{(a)} \; + \;
 \; i\; {e \over c}  \; A^{(a)}\; \Psi _{(a)}  =
\; i\; {e \over 2cM }  \;
 F ^{(a)(b)} \; \Psi _{(a)(b)}    \; .
\label{3.3'}
\end{eqnarray}

\noindent The Coulomb field  $ A_{0 } = e  /r \; , \; F_{r0 } = -
e/ r^{2}$ in the tetrad description looks
\begin{eqnarray}
A^{(0)} = e^{(0) 0  } A_{0 } =  {e \over r}  \; ,  \qquad
F^{30} =  e^{(3)r }  e^{(0)0 } \; F_{r0}  =
  -  { e \over r^{2}} \; .
\label{3.4'}
\end{eqnarray}

\noindent Beside, after simple calculation we get
\begin{eqnarray}
e^{(0)\alpha }_{\;\;\; ;\alpha } = 0 \; , \qquad  e^{(1)\alpha
}_{\;\;\; ;\alpha } = - {\cos \theta \over r \sin \theta } \;
,\qquad e^{(2)\alpha }_{\;\;\; ;\alpha } = 0\; , \qquad
e^{(3)A}_{\;\;\; ;\alpha } = - {2 \over  r}  \; . \nonumber
\end{eqnarray}

\noindent The components functions  $\Psi _{(a)}$ and   $\Psi _{(a)(b)}$ in  (\ref{3.3'})
can be related with components of the D-K-P column as follows
 (transition between cyclic and Cartesian representations; $c \equiv 1/ \sqrt{2}$)
\begin{eqnarray}
\left | \begin{array}{l}
\Psi_{(0)} \\ \Psi_{(1)} \\ \Psi_{(2)} \\ \Psi_{(3)} \\
         \Psi_{(0)(1)} \\ \Psi_{(0)(2)} \\ \Psi_{(0)(3)} \\
         \Psi_{(2)(3)} \\ \Psi_{(3)(1)} \\ \Psi_{(1)(2)}
\end{array} \right | =
\left | \begin{array}{cccccccccc}
1      & 0 & 0 &  0      & 0 & 0 & 0      & 0 & 0 &  0 \\
0      &-c & 0 & +c      & 0 & 0 & 0      & 0 & 0 &  0 \\
0      &-ic& 0 &-ic      & 0 & 0 & 0      & 0 & 0 &  0 \\
0      & 0 & 1 &  0      & 0 & 0 & 0      & 0 & 0 &  0 \\
0      & 0 & 0 & 0       &-c & 0 & +c     & 0 & 0 &  0 \\
0      & 0 & 0 & 0       &-ic& 0 &-ic     & 0 & 0 &  0 \\
0      & 0 & 0 & 0       & 0 & 1 & 0      & 0 & 0 &  0 \\
0      & 0 & 0 & 0       & 0 & 0 & 0      &-c & 0 & +c \\
0      & 0 & 0 & 0       & 0 & 0 & 0      &-ic& 0 &-ic \\
0      & 0 & 0 & 0       & 0 & 0 & 0      & 0 & 1 &  0
\end{array} \right | =
\left | \begin{array}{l}
    \Phi_{0} \;D_{0}    \\  \Phi_{1} \;D_{-1} \\  \Phi_{2} \;D_{0} \\
    \Phi_{3} \; D_{+1} \\  E_{1} \; D_{-1} \\  E_{2} D_{0}   \\
    E_{3} \; D_{+1} \\  H_{1} \;D_{-1}  \\  H_{2} D_{0}   \\
    H_{3}\;D_{+1}
\end{array} \right |     \; .
\nonumber
\end{eqnarray}

\noindent We need only  $\Phi _{0}, \;\Psi _{(1)}
, \;\Psi _{(2)} , \;\Psi _{(3)} , \;\Psi _{(0)(3)}:$
\begin{eqnarray}
\Psi _{(1)} = \; e^{-iEt/ \hbar} {1 \over \sqrt{2}} \; (-
\Phi_{1} \;D_{ -1} + \Phi_{3}  \; D_{ +1} ) \; ,
\nonumber
\\
\Psi _{(2)} = \; e^{-iE t / \hbar}  {i \over \sqrt{2}} \; (-
\Phi_{1} \;D_{ -1}   - \Phi_{3} \; D_{ +1} ) \;. \nonumber
\\
\Psi _{(0)} = \; e^{-iE  t / \hbar}  \Phi_{0} \; D_{0 } \; , \qquad
\Psi _{(3)} = \; e^{-iEt / \hbar}  \Phi_{2}  \; D_{0 } \; ,
\nonumber
\\
\Psi _{(0)(3)} = e^{-i E t /\hbar } \; E_{2}\; D_{0 }     \; .
\label{3.6'}
\end{eqnarray}

\noindent With the help of  (\ref{3.6'}), eq.    (\ref{3.3'}) gives
\begin{eqnarray}
 {1 \over \sqrt{2}} \;  r \; \Phi_{1} \; (
\partial _{\theta } \; D_{ -1}  - {M   -1 \cos
\theta \over \sin \theta} \;  D_{-1} )  -
\nonumber
\\
- {1 \over \sqrt{2}} \; r \;  \Phi_{3}\; ( \partial _{\theta } \;
D_{ +1} +  {M   +1  \cos \theta ) \over \sin \theta} \;  D _{ +1})
+ \; \nonumber
\\
+
 D_{0 } \;  ( - {2 \over r} \; \Phi_{2} \; - \;
i \epsilon  \; \Phi_{0} \; - \; {d \over dr}\;
\Phi_{2} )
 =
  i \; { \alpha  \over 2M r^{2}} \; E_{2} D_{0}      \; .
\nonumber
\end{eqnarray}

\noindent Now, by taking into account the recurrent formulas
\cite{Varshalovich-Moskalev-Hersonskiy-1975}
\begin{eqnarray}
\partial _{\theta } \; D_{ -1} -
{M  -1 \cos \theta \over \sin \theta } \; D_{-1} = - \sqrt{(j  +1)
j} \; D_{0 } \; , \nonumber
\\
\partial _{\theta } \; D_{+1} -
{M +  1\cos \theta \over \sin \theta } \; D_{ +1} = - \sqrt{(j+ 1)
j } \; D_{0  }    \; , \nonumber
\end{eqnarray}

\noindent we arrive at a the Lorentz condition in radial form
\begin{eqnarray}
 - i \; \epsilon  \; \Phi_{0}  -
\; ({d \over dr} + {2 \over r})\; \Phi_{2}  -  {\nu \over r } \; (
\Phi_{1} +  \Phi_{3})  =
   { i\alpha   \over 2 M r^{2}} \;  E_{2}  \; .
\label{3.7'}
\end{eqnarray}

For states with parity $P= (-1)^{j+1}$,  this condition is satisfied identically.
For states with parity   $P= (-1)^{j}$ it gives
\begin{eqnarray}
 - i \; \epsilon  \; \Phi_{0}  -
\; ({d \over dr} + {2 \over r})\; \Phi_{2}  -  {2\nu \over r } \;
\Phi_{1}   =
  { i\alpha    \over 2 M r^{2}} \;  E_{2}  \; .
\label{3.8''}
\end{eqnarray}

From eq.  (\ref{3.8''})  a very important relationship can be established.
To this end ,  from  eq. (\ref{3.8''})  let us exclude  $\Phi_{2}$  with the help of the third equation
in   (\ref{1.12})
\begin{eqnarray}
  i \; \epsilon M \; \Phi_{0}  +
 i  ( \epsilon + {\alpha \over r})   ({d \over dr} + {2 \over r}) E_{2} - {2i\nu \over r}
({d \over dr} + {1 \over r})   H_{1}  +   {2\nu  M \over r } \;
\Phi_{1}   =
   i \; { \alpha   \over 2  r^{2}} \;  E_{2}  \; .
\nonumber
\end{eqnarray}

\noindent Let us transform the second and third terms with the help of
 1-st and 2-nd equations in  (\ref{1.12}) -- it results in
 \begin{eqnarray}
   E_{2} =  -2   M r \;       \Phi_{0}          \; .
 \label{3.9'}
 \end{eqnarray}

\section{ The  main function  of the first type,  reducing
the problem to the confluent Heun equation
 }

Let  us examine  eqs. (\ref{1.12}),
now with the use od the Lorentz condition  (\ref{3.8''})  and its consequence  (\ref{3.9'}).
From the 1-st eq. in  (\ref{1.12}), we  produce
\begin{eqnarray}
 E_{1} = {Mr
\over 2 \nu } (5 + 2r {d\over dr}) \Phi_{0} \; . \label{4.2}
 \end{eqnarray}

With the help of (\ref{3.9'}), the fourth eq. in  (\ref{1.12}) gives
\begin{eqnarray}
 \Phi_{1} = {- i \over \epsilon + \alpha  / r} \left  ( {\nu \over r}  -
{ 5 M^{2}  \over 2 \nu } r - {r^{2}M^{2} \over  \nu }    {d\over
dr}  \right ) \Phi_{0}  \; . \label{4.3}
\end{eqnarray}

With the help of (\ref{3.9'}), the fifth eq. in  (\ref{1.12}) gives
\begin{eqnarray}
\Phi_{2} = {i \over \epsilon + \alpha  / r} \; \left ( {d \over dr}
-  2M^{2}r  \right ) \; \Phi_{0}  \; , \label{4.4}
\end{eqnarray}

The sixth eq. in  (\ref{1.12}) provides us with the representation for
$H_{1}$ in term of $\Phi_{0}$ by means of second order differential operator
\begin{eqnarray}
 - M H_{1} = \left [
  ({d \over dr}  +{1\over r})
 {1\over \epsilon +  \alpha  / r} \left  ( {\nu \over r}  -
{ 5 M^{2}  \over 2 \nu } r - {r^{2}M^{2} \over  \nu }    {d\over
dr}  \right )   -
  {\nu \over r}   {1 \over \epsilon + \alpha / r}  \left ( {d \over dr}  -  2M^{2}r  \right ) \right ]  \Phi_{0} .
\label{4.5}
\end{eqnarray}

\noindent Take note that from the third eq. in (\ref{1.12}) one can obtain another
representation for
$H_{1}$ that uses only the first order operator
\begin{eqnarray}
H_{1} =   - {Mr  \over 2 \nu} \; {1 \over  \epsilon + \alpha  / r }
\; \left [ {d \over dr } + 2r \; [ \; (\epsilon +{ \alpha  \over
r})^{2} - M^{2} \; ] \right ] \Phi_{0} \; . \label{4.5'}
\end{eqnarray}

\noindent
These two representations for  $H_{1}$ must be consistent  with each other.

Now, turning to the Lorentz condition
\begin{eqnarray}
 - i \; \epsilon  \; \Phi_{0}  -
\; ({d \over dr} + {2 \over r})\; \Phi_{2}  -  {2\nu \over r } \;
\Phi_{1}   =
  i \; { \alpha    \over 2 M r^{2}} \;  E_{2}  \; .
\nonumber
\label{4.6}
\end{eqnarray}

\noindent one can readily derive a second order differential equation for
$\Phi_{0}$:
\begin{eqnarray}
{d^{2} \Phi_{0} \over dr^{2}} + {1 \over r} \left ( 3 - {\epsilon
\over  \epsilon +  \alpha  / r}  \right ) {d  \Phi_{0} \over dr} +
 \left ( \epsilon^{2} - {\alpha^{2} \over r^{2}} -3M^{2} + 2M^{2} {\epsilon \over \epsilon + \alpha / r } -
 { 2\nu^{2} \over r^{2} } \right ) \Phi_{0} = 0 \; .
\label{4.7}
\end{eqnarray}

\noindent
The function $\Phi_{0}$ will be termed as  main function, 5 remaining ones
 $\Phi_{1}, \Phi_{2},  E_{1}, H_{1}$ are expressed through it
$$
   E_{2} =  -2   M r \;       \Phi_{0}          \; ,
\eqno1
$$
$$
E_{1} = {Mr \over 2 \nu } (5 + 2r {d\over dr}) \Phi_{0}\; , \eqno2
$$
$$
\Phi_{1} = {- i \over \epsilon + \alpha  / r} \left  ( {\nu \over r}
- { 5 M^{2}  \over 2 \nu } r - {r^{2}M^{2} \over  \nu }    {d\over
dr}  \right ) \Phi_{0}\; , \eqno3
$$
$$
\Phi_{2} = {i \over \epsilon +  \alpha / r} \; \left ( {d \over dr}
-  2M^{2}r  \right ) \; \Phi_{0} \; , \eqno4
$$
$$
H_{1} =   - {Mr  \over 2 \nu} \; {1 \over  \epsilon +  \alpha  / r }
\; \left [ {d \over dr } + 2r \; [ \; (\epsilon +{ \alpha \over
r})^{2} - M^{2} \; ] \right ] \Phi_{0} \; . \eqno5
$$
\begin{eqnarray}
\label{aux}
\end{eqnarray}

Changing the variable
\begin{eqnarray}
x = - {\epsilon  \over \alpha  } \; r\; < 0 \;  , \qquad r = - { \alpha  \over
\epsilon } \;  x \;  ;
\nonumber
\end{eqnarray}

\noindent eq.  (\ref{4.7}) is reduced to the form
\begin{eqnarray}
 {d^{2} \Phi_{0} \over dx^{2}} +   \left  ( {3 \over x } - {1  \over  x- 1 }  \right ) {d  \Phi_{0} \over dx} +
 \left ( \alpha^{2} - \Lambda^{2} -  {\alpha^{2} + 2 \nu^{2}  \over x^{2}}  +   {2 \Lambda^{2}  \over x- 1  }
 \right ) \Phi_{0} = 0 \; ;
\label{4.9}
\end{eqnarray}

\noindent  where dimensionless parameters were used
\begin{eqnarray}
\Lambda^{2} =   \alpha^{2} \; \lambda^{2} \; , \qquad  \lambda  ={mc^{2} \over
E} > 1 \; .
\nonumber
\end{eqnarray}

Let us consider behavior of the main function  near the point  $x=0$:
\begin{eqnarray}
 {d^{2} \Phi_{0} \over dx^{2}} +   {3 \over x }  {d  \Phi_{0} \over dx}
    - { \alpha^{2} +2\nu^{2} \over x^{2}}     \;  \Phi_{0} = 0 \; ,
    \nonumber
   \\
      \Phi_{0} \sim \mbox{const} \;  x^{A} , \qquad A(A-1) +3A - \alpha^{2} - 2\nu^{2} = 0 \; ,\nonumber
   \\
   A = -1 - \sqrt{1 + \alpha^{2} + 2\nu^{2} } \; ,  \;  A = -1 + \sqrt{1 + \alpha^{2} + 2\nu^{2} } \; ;
   \label{4.10}
\end{eqnarray}

\noindent to bound states there correspond positive values of $A$.
In the region near $x=+\infty$,  the main equation gives
\begin{eqnarray}
 {d^{2} \Phi_{0} \over dx^{2}} +   {2 \over x } \;
{d  \Phi_{0} \over dx} +
 \left ( \alpha^{2}  -  \Lambda^{2}     \right ) \Phi_{0} = 0 \; ,
 \qquad
  \Phi_{0} = e^{+  \sqrt{\Lambda^{2} - \alpha^{2}} \; x } =
  e^{-   \sqrt{ m^{2}c^{4} - E^{2} }\; r / \hbar c  } \;  ;
\label{4.11}
\end{eqnarray}

\noindent to bound states there correspond  solutions vanishing at infinity.

Now, let us introduce substitution
$
\Phi_{0}(x)  = x^{A} e^{Bx}\; f (x)$,
eq.  (\ref{4.9}) gives
\begin{eqnarray}
 {d^{2} f \over dx^{2}} +
\left[ 2B +{2A+3 \over x }+ {1  \over  1-x } \right] {d  f \over
dx} +
 \left [ B^{2}+ \alpha^{2}  -  \Lambda^{2} +{2AB+A+3B\over x}+
 \right.
 \nonumber
 \\
 \left.
 {A(A-1)+3A- \alpha^{2}-2\nu^{2}  \over x^{2}}+
  {A+B -2 \Lambda^{2}  \over 1-x}  \right ] \; f (x) = 0 \; .
\label{4.14}
\end{eqnarray}

\noindent
With restrictions on  $A$ and   $B$:
\begin{eqnarray}
A(A-1)+3A- \alpha^{2}-2\nu^{2}=0 \;\; \;  \Longrightarrow \;\;\;
A=-1 + \sqrt{1+2\nu^{2}+ \alpha^{2}}\,; \nonumber
\\[3mm]
B^{2}+ \alpha^{2}  - \Lambda ^{2} =0  \qquad   \Longrightarrow
\qquad
  B= + \sqrt{ \Lambda^{2} - \alpha^{2}}  \,,
\label{4.14'}
\end{eqnarray}

\noindent eq.   (\ref{4.14}) takes the form
\begin{eqnarray}
 {d^{2} f \over dx^{2}} +
\left[ 2B +{2A+3 \over x } - {1  \over  x- 1 } \right] {d  f \over
dx} +
\nonumber
\\
+
 \left [ {2AB+A+3B\over x}+{A+B -2\Lambda^{2}  \over 1-x}  \right ]f = 0 \; .
\label{4.15}
\end{eqnarray}

It can be recognized as the confluent Heun's  equation
\begin{eqnarray}
f = f(a,b,c, d, h ; \; z) \;, \qquad {d^{2} f  \over dx^{2}}  +
\left ( a + {b+1 \over x } + {c+1 \over x-1} \right ){d f \over d
x} - \nonumber
\\
- {[-2  d +a (- b-c-2)]x +a (1+b) +b (-1-c)-c -2 h \over 2 x(x-1)
} \; f =0
\label{4.16'}
\end{eqnarray}

with parameters given by
\begin{eqnarray}
a =  +2 \sqrt{ \Lambda^{2} - \alpha^{2}}\; , \qquad b  =  +2
\sqrt{1+2\nu^{2}+ \alpha^{2}}\;, \nonumber
\\
c  =-2 \; , \qquad d = 2\Lambda^{2} \; , \qquad h = + 2 \; .
\label{4.17'}
\end{eqnarray}

The known condition for polynomial solutions is
\begin{eqnarray}
d=-a\left(n+{b+c+2\over 2}\right)\,,
\label{4.18}
\end{eqnarray}

\noindent it  gives  the following quantization rule
\begin{eqnarray}
{\Lambda^{4}\over
\Lambda^{2}-\alpha^{2}}=(n+\sqrt{1+2\nu^{2}+\alpha^{2}})^{2}\,.
\label{4.19}
\end{eqnarray}

\noindent
Its physical solution is
\begin{eqnarray}
E^{2}  =   m^{2}c^{4} \; {2\alpha^{2}  \over   N^{2}   -
\sqrt{N^{4} - 4\alpha^{2} N^{2} } } \; .
\label{4.20}
\end{eqnarray}

\noindent
When $N$ increases to infinity, we get
\begin{eqnarray}
N \rightarrow \infty , \qquad E^{2}  =   m^{2}c^{4} \;
{2\alpha^{2}  \over   N^{2}   -  N^{2} \sqrt{ 1 - 4\alpha^{2}
N^{-2} } } \approx m^{2}c^{4} \; {2\alpha^{2} \over 2 \alpha ^{2}
} =  m^{2}c^{4} \; .
\label{4.21}
\end{eqnarray}

To obtain a non-relativistic limit, one must  impose  special restriction
\begin{eqnarray}
N = n+\sqrt{1+2\nu^{2}+\alpha^{2}}   =  n+\sqrt{1+j(j+1)  +
\alpha^{2}} \approx n + j
\label{4.22}
\end{eqnarray}

\noindent  which correlates  with  the known non-relativistic procedure
\begin{eqnarray}
m + \epsilon + {\alpha \over r}  \approx m + \epsilon \; .
\nonumber
\end{eqnarray}

\noindent  Taking into account  eq. (\ref{4.22}), one  can we derive
\begin{eqnarray}
E^{2} = m^{2}c^{4} \;{1 \over 2}  \; (1   +   \sqrt{1 -
{4\alpha^{2} \over N^{2} } } ) \approx m^{2}c^{4} (1 - {\alpha^{2}
\over N^{2}} ) \, \qquad N \approx n + j \; ,
\nonumber
\end{eqnarray}

\noindent
that is
\begin{eqnarray}
E =    mc^{2}
(1 - {\alpha^{2} \over  2 N^{2}} )= mc^{2} + E' \; .
\label{4.23}
\end{eqnarray}

\noindent
Thus, the  non-relativistic  energy levels are  given by
\begin{eqnarray}
E' =-  {\alpha^{2} mc^{2} \over  2 N^{2}}  = -{ m e^{4} \over  \hbar^{2} N^{2}} \;,
\label{4.24}
\end{eqnarray}

\noindent which coincides with  the known exact result (see below).

\section{ The main radial function of the second type,
reducing the problem to another  second order differential  equation
}

From general considerations, we may expect two linearly independent solutions for
6-equation system for state with parity $P=(-1)^{j}$. The above equation
 (\ref{4.7}) provides us  with  only one class of these, what make us look for  another
 class (may with with some different main function).

In this connection, let us turn back to eq. (\ref{4.5'})  multiplied by  $-M$  and  compare it with eq.(\ref{4.5}),
from that it follows a second order differential equation $\Phi_{0}$,  different from  (\ref{4.7}):
\begin{eqnarray}
{d^{2}\Phi_{0}\over dr^{2}}+{1\over r}\,(6+{ \alpha \over r
(\epsilon+ \alpha /r)})\,{d\Phi_{0}\over dr}+ \nonumber
\\
+\left[\epsilon^{2}-M^{2}+{2\epsilon^{2} \alpha \over \epsilon
r+ \alpha }- { \alpha  \nu^{2}\over r^{4}M^{2}(\epsilon r+ \alpha )}-
\right. \nonumber
\\
\left. - {1\over 2}{\alpha (-15+4\nu^{2}-2 \alpha^{2})\over
r^{2}(\epsilon r+ \alpha )}-{\epsilon(-5+2\nu^{2}-3 \alpha ^{2})\over r
(\epsilon r+ \alpha )}\right]\Phi_{0}=0\,. \label{6.1}
\end{eqnarray}

\noindent In  the variable  $x$, it reads
\begin{eqnarray}
{ d^{2} \over dx^{2}} +  {1 \over x} \left ( 6 -{x \over x -1}
\right ) {d \over dx} +  \left [ (1 - \lambda^{2}) \alpha^{2}  -{2 \alpha^{2}  \over  x
-1} +
 {   \nu^{2} \over  \alpha^{2} \lambda^{2} x^{4} (x-1)}
+ \right. \nonumber
\\
\left. + {(-15 + 4\nu^{2} - 2 \alpha^{2}) \over  2x^{2} (x-1) } -
{ (5+2\nu^{2} - 3\alpha^{2}) \over   x (x-1)} \right ] \Phi_{0}=0
\; . \label{6.2}
\end{eqnarray}

By means of the coordinate  transformation  $y = x^{-1}$, eq.
(\ref{6.2}) becomes
\begin{eqnarray}
{d^{2}\Phi_{0}\over dy^{2}}+{4y-3\over y(1-y)}\,{d\Phi_{0}\over
dy}+   \left [ {(1 - \lambda^{2}) \alpha^{2}\over y^{4}}  -{2
\alpha^{2}  \over  y^{3}(1-y)} +
 {   \nu^{2}y \over  \alpha^{2} \lambda^{2} (1-y)}
- \right. \nonumber
\\
\left. - {(15 - 4\nu^{2} + 2 \alpha^{2}) \over  2y (1-y) } - {
(5+2\nu^{2} - 3\alpha^{2}) \over   y^{2} (1-y)} \right ]
\Phi_{0}=0 \; . \label{6.8}
\end{eqnarray}

\noindent
Both differential equations,  (\ref{6.2}) and  (\ref{6.8}), are
very complex. We might expect that  they can describe some third class of solutions,
however  any proof of this  does not exist now.

\section{ Non-relativistic limit, exact energy spectrum
}

In treating this  point we will use results on non-relativistic limit for a vector particle according to
\cite{Bogush-Kisel-Tokarevskaya-Red'kov-2002, Bogush-Kisel-Tokarevskaya-Red'kov-2007},
general treatment of the problem of the wave equations for arbitrary spin particle see in
\cite{Levy-Leblond-1967}, \cite{1974-Fushchych, 1976-Fushchych};
also see in
\cite{Moshin-Tomazelli-2008}.

First, let us consider the simpler system  (\ref{8.1}) for states with parity
 $P=(-1)^{j+1}$:
\begin{eqnarray}
\underline{P=(-1)^{j+1}} , \qquad + i (\epsilon + {\alpha \over
r}) E_{1} +  i ({d \over dr} +{ 1 \over r}) H_{1} +
 i {\nu \over r} H_{2}
= M \Phi _{1} \; , \nonumber
\\
-i (\epsilon + {\alpha \over r})  \Phi_{1}  = M E_{1} \; , \;\; -i
({d \over dr} + {1 \over r}) \Phi_{1}   = M H_{1} \; , \;\; 2i {
\nu \over r }  \Phi_{1} = M H_{2}  \; .
\label{8.1}
\end{eqnarray}

Here the $H_{1},H_{2}$ represent non-dynamical variables, excluding them we obtain
\begin{eqnarray}
+ i (\epsilon + {\alpha \over r}) E_{1} +    {1 \over M}\; ({d
\over dr} +{ 1 \over r})^{2}
 \; \Phi_{1}   -
{ 2  \nu^{2} \over M r^{2} } \; \Phi_{1} = M \Phi _{1} \; ,\qquad
-i (\epsilon + {\alpha \over r})  \Phi_{1}  = M E_{1} \; .
\label{8.4}
\end{eqnarray}

\noindent Now we should make special substitution, introducing a big  and small constituents
( $B_{1}(r)$   and  $M_{1}(r)$ respectively)
\begin{eqnarray}
\Phi_{1} =  B_{1} + M_{1}  \; , \qquad  iE_{1}  =   B_{1} - M_{1}
\; ; \label{8.5}
\end{eqnarray}

\noindent correspondingly eqs.  (\ref{8.4}) take the form
\begin{eqnarray}
(\epsilon + {\alpha \over r})(B_{1} - M_{1})  +
 {1 \over M} ({d \over dr} +{ 1 \over r})^{2}
  (B_{1} + M_{1} )
   -
{ 2  \nu^{2} \over M r^{2} }  (B_{1} + M_{1} ) = M (B_{1} +
M_{1} ) \; , \nonumber
\\
(\epsilon + {\alpha \over r}) (B_{1} + M_{1} )  = M ( B_{1} -
M_{1} )  \; . \label{8.6}
\end{eqnarray}

\noindent Summing and subtracting these two we get
\begin{eqnarray}
2 (\epsilon + {\alpha \over r}) B_{1}  +
 {1 \over M}\; ({d \over dr} +{ 1 \over r})^{2}
 \;  (B_{1} + M_{1} )  -
{ 2  \nu^{2} \over M r^{2} } \; (B_{1} + M_{1} ) = 2 M B_{1}  \; ,
\label{8.7a} \nonumber
\\
- 2 (\epsilon + {\alpha \over r}) M_{1}  +
 {1 \over M}\; ({d \over dr} +{ 1 \over r})^{2}
 \;  (B_{1} + M_{1} )  -
{ 2  \nu^{2} \over M r^{2} } \; (B_{1} + M_{1} ) = 2 M M_{1}  \; .
\label{8.7b}
\end{eqnarray}

\noindent Now we should separate  a rest energy by  a formal change $
\epsilon  \Longrightarrow  \epsilon + M $; which results in
\begin{eqnarray}
2 (\epsilon + {\alpha \over r}) B_{1}  +
 {1 \over M}\; ({d \over dr} +{ 1 \over r})^{2}
 \;  (B_{1} + M_{1} )  -
{ 2  \nu^{2} \over M r^{2} } \; (B_{1} + M_{1} ) = 0 \; ,
\nonumber
\\
- 2 (\epsilon + {\alpha \over r}) M_{1}  +
 {1 \over M}\; ({d \over dr} +{ 1 \over r})^{2}
 \;  (B_{1} + M_{1} )  -
{ 2  \nu^{2} \over M r^{2} } \; (B_{1} + M_{1} ) = 4 M M_{1}    \;
. \nonumber
\end{eqnarray}

\noindent Thus, we produce equation for a big  $B_{1}(r)$   and small $M_{1}(r)$ components
\begin{eqnarray}
2 (\epsilon + {\alpha \over r} ) B_{1}  +
 {1 \over M}\; ({d \over dr} +{ 1 \over r})^{2}
 \;  B_{1}   -
{ j(j+1)  \over M r^{2} } \; B_{1} = 0  \; , \label{8.9a}
\\
({d \over dr} +{1 \over r})^{2} B_{1} - {j(j+1) \over r^{2} }
B_{1} = 4 M^{2} M_{1} \; . \label{8.9b}
\end{eqnarray}

\noindent Equation for the big component can be written as  Schr\"{o}dinger equation for a scalar particle
\begin{eqnarray}
\left [  {d^{2} \over dr^{2} } + {2 \over r} {d \over dr } + 2 M (
\epsilon + {\alpha \over r} ) - { j(j+1)  \over  r^{2} } \right ]
B_{1} = 0   \; . \label{8.10}
\end{eqnarray}

\noindent Corresponding non-relativistic 3-dimensional wave function
for states with parity  $P = (-1)^{j+1}$ is
\begin{eqnarray}
 P = (-1)^{j+1}\; , \qquad \Psi = e^{-iEt/ \hbar} \left | \begin{array}{c}
+ (\Phi_{1} + i E_{1})\;  D_{-1} \\
0 \\
-(\Phi_{1} + i E_{1})  \; D_{+1}
\end{array} \right |  .
\label{non-relat'}
\end{eqnarray}

Now let us consider  radial equations for  states with  opposite parity $P=(-1)^{j}$
\begin{eqnarray}
 - ({d \over dr} +{2\over r}) E_{2} - 2 {\nu \over r} E_{1} = M \Phi_{0}\; ,
\qquad
+ i (\epsilon + {\alpha \over r}) E_{1} + i ( {d \over dr} + {1
\over r}) H_{1} = M \Phi_{1} \; , \nonumber
\\
+ i (\epsilon  + {\alpha \over r}) E_{2} -2i {\nu \over r}  H_{1}
= M \Phi_{2} \; , \qquad
-i (\epsilon + {\alpha \over r})  \Phi_{1} + {\nu \over r}
\Phi_{0} = M E_{1} \; , \; \nonumber
\\
-i (\epsilon + {\alpha \over r}) \Phi_{2}  - {d \over dr} \Phi_{0}
= M E_{2} \; , \qquad
-i ({d \over dr} +{1\over r}) \Phi_{1} -i {\nu \over r} \Phi_{2} =
M H_{1} \; . \label{8.2}
\end{eqnarray}

Among  four dynamical  function $\Phi_{1}, \Phi_{2}, E_{1} , E_{2}$
separation of big and small constituents is performed as follows
\begin{eqnarray}
\Phi_{1} = B_{1} +  M_{1} \; , \qquad \Phi_{2}
= B_{2} +  M_{2} \; , \qquad
i E_{1} = B_{1} -  M_{1} \; , \qquad  i E_{2}
= B_{2} -  M_{2} \; ; \; \label{big-small}
\end{eqnarray}

\noindent the non-relativistic 3-dimensional wave function
for states with parity  $P = (-1)^{j}$ is defined
\begin{eqnarray}
 P = (-1)^{j}\; , \qquad \Psi = e^{-iEt/ \hbar} \left | \begin{array}{l}
 (\Phi_{1} + i E_{1})\;  D_{-1} \\
(\Phi_{2} + i E_{2})  \; D_{0} \\
(\Phi_{1} + i E_{1})  \; D_{+1}
\end{array} \right |  .
\label{non-relat}
\end{eqnarray}

 Excluding from  (\ref{8.2}) non-dynamical variables
 $\Phi_{0}, H_{1}$, we obtain the system (the rest energy is taken away as well:  $ \epsilon
\Longrightarrow \epsilon + M  $)
\begin{eqnarray}
i (\epsilon + M + {\alpha \over r}) \; E_{1} + {1 \over M}\; ({d
\over dr} + {1 \over r})\; [\;({d \over dr} +{1\over r}) \Phi_{1}
+ {\nu \over r} \Phi_{2} \; ] \; = M \Phi_{1} \; , \nonumber
\\
i (\epsilon + M+ {\alpha \over r}) E_{2} - {2 \nu \over M r} \; [
\; ({d \over dr} +{1\over r}) \Phi_{1} + {\nu \over r} \Phi_{2} \;
] \; = M \Phi_{2} \; , \nonumber
\\
-i (\epsilon + M + {\alpha \over r}) \Phi_{1} + { \nu \over  M r }
\; [ \; -({d \over dr} + {2 \over r}) E_{2} - {2\nu \over r} E_{1}
\; ] = M E_{1} \; , \nonumber
\\
-i (\epsilon +M +  {\alpha \over r}) \Phi_{2} - { 1 \over  M } {d
\over dr} \; [ \; -({d \over dr} + {2 \over r}) E_{2} - {2\nu
\over r} E_{1} \; ] = M E_{2} \; .
\label{8.11}
\end{eqnarray}

\noindent Taking into account  (\ref{big-small}), transform  (\ref{8.11}) into
\begin{eqnarray}
(\epsilon + M + {\alpha \over r })( B_{1} -M_{1}) + {1 \over M} (
{d \over dr }+ {1 \over r } ) ^{2} (B_{1} +M_{1}) +
\nonumber
\\
+ { \nu \over M
} ( {d \over dr } +  {1 \over r}) {1 \over r} (B_{2} + M_{2}) = M
(B_{1}  + M_{1})   , \nonumber
\\
(\epsilon + M + {\alpha \over r} ) (B_{2} - M_{2})  - {2 \nu \over
M r } ({d \over dr} + {1 \over r}) (B_{1}  +M_{1}) -
\nonumber
\\
- {2 \nu^{2}
\over M r^{2}} (B_{2} + M_{2}) = M (B_{2} +M_{2}) , \nonumber
\\
(\epsilon + M + {\alpha \over r}) (B_{1}+M_{1})  - { \nu \over mr}
({d \over dr} + {2 \over r}) (B_{2} -M_{2}) -
\nonumber
\\
- {2\nu^{2} \over
Mr^{2}} (B_{1} -M_{1})  = M (B_{1} - M_{1})  , \nonumber
\\
(\epsilon + M + {\alpha \over r}) (B_{2} + M_{2})  + {1 \over M}
{d \over dr} ({d \over dr} +
\nonumber
\\
+ {2 \over r}) (B_{2} -M_{2}) + {2\nu
\over M}{d \over dr}{1 \over r}( B_{1}-M_{1})  = M( B_{2} -M_{2})
. \nonumber \label{6.12}
\end{eqnarray}

\noindent From whence, we get
\begin{eqnarray}
(\epsilon+ {\alpha \over r })( B_{1} -M_{1}) + {1 \over M} ( {d
\over dr }+ {1 \over r } ) ^{2} (B_{1} +M_{1}) +
\nonumber
\\
+  { \nu \over M }
( {d \over dr } +  {1 \over r}) {1 \over r} (B_{2} + M_{2}) = + 2
M  \; M_{1}   , \nonumber
\\
(\epsilon +  {\alpha \over r}) (B_{1}+M_{1})  - { \nu \over M r}
({d \over dr} + {2 \over r}) (B_{2} -M_{2}) -
\nonumber
\\
-  {2\nu^{2} \over M
r^{2}} (B_{1} -M_{1})  = -2 M  M_{1}  , \nonumber
\\
(\epsilon  + {\alpha \over r} ) (B_{2} - M_{2})  - {2 \nu \over M
r } ({d \over dr} + {1 \over r}) (B_{1}  +M_{1}) -
\nonumber
\\
-
{2 \nu^{2}
\over M r^{2}} (B_{2} + M_{2}) = + 2 M \; M_{2} , \nonumber
\\
(\epsilon  + {\alpha \over r}) (B_{2} + M_{2})  + {1 \over M} {d
\over dr} ({d \over dr} +
\nonumber
 \\
 + {2 \over r}) (B_{2} -M_{2}) + {2\nu
\over M}{d \over dr}{1 \over r}( B_{1}-M_{1})  =  -2 M \; M_{2}  .
\nonumber \label{6.13}
\end{eqnarray}

\noindent Now summing and subtracting equation within the first couple,
and doing the same within second couple, we arrive at
\begin{eqnarray}
(\epsilon+ {\alpha \over r })( B_{1} -M_{1}) + {1 \over M} ( {d
\over dr }+ {1 \over r } ) ^{2} (B_{1} +M_{1}) +  { \nu \over M }
( {d \over dr } +  {1 \over r}) {1 \over r} (B_{2} + M_{2}) +
\nonumber
\\
+ (\epsilon +  {\alpha \over r}) (B_{1}+M_{1})  - { \nu \over M r}
({d \over dr} + {2 \over r}) (B_{2} -M_{2}) - {2\nu^{2} \over M
r^{2}} (B_{1} -M_{1})  =  0  \;, \nonumber
\end{eqnarray}

\begin{eqnarray}
(\epsilon+ {\alpha \over r })( B_{1} -M_{1}) + {1 \over M} ( {d
\over dr }+ {1 \over r } ) ^{2} (B_{1} +M_{1}) +  { \nu \over M }
( {d \over dr } +  {1 \over r}) {1 \over r} (B_{2} + M_{2}) -
\nonumber
\\
- (\epsilon +  {\alpha \over r}) (B_{1}+M_{1})  + { \nu \over M r}
({d \over dr} + {2 \over r}) (B_{2} -M_{2}) +
\nonumber
\\
+ {2\nu^{2} \over M
r^{2}} (B_{1} -M_{1})  =  + 4 M  \; M_{1}  \; , \nonumber
\end{eqnarray}
\begin{eqnarray}
(\epsilon  + {\alpha \over r} ) (B_{2} - M_{2})  - {2 \nu \over M
r } ({d \over dr} + {1 \over r}) (B_{1}  +M_{1}) - {2 \nu^{2}
\over M r^{2}} (B_{2} + M_{2}) + \nonumber
\\
+ (\epsilon  + {\alpha \over r}) (B_{2} + M_{2})  + {1 \over M} {d
\over dr} ({d \over dr} + {2 \over r}) (B_{2} -M_{2}) + {2\nu
\over M}{d \over dr}{1 \over r}( B_{1}-M_{1})  =  0  \;, \nonumber
\end{eqnarray}
\begin{eqnarray}
(\epsilon  + {\alpha \over r} ) (B_{2} - M_{2})  - {2 \nu \over M
r } ({d \over dr} + {1 \over r}) (B_{1}  +M_{1}) - {2 \nu^{2}
\over M r^{2}} (B_{2} + M_{2}) - \nonumber
\\
- (\epsilon  + {\alpha \over r}) (B_{2} + M_{2})  - {1 \over M} {d
\over dr} ({d \over dr} + {2 \over r}) (B_{2} -M_{2}) -
\nonumber
\\- {2\nu
\over M}{d \over dr}{1 \over r}( B_{1}-M_{1})  = + 4 M \; M_{2}\; .
\nonumber
\end{eqnarray}

Now, using the same method as in
\cite{Bogush-Kisel-Tokarevskaya-Red'kov-2002, Bogush-Kisel-Tokarevskaya-Red'kov-2007}
 (consider $B_{1}, B_{2} $ as big, and   $M_{1},M_{2}$ as small),
we arrive at two equations for big components, and  two equations defining small components through big ones:
\begin{eqnarray}
{1 \over M} ( {d \over dr }+ {1 \over r } ) ^{2} B_{1}  +  { \nu
\over M } ( {d \over dr } +  {1 \over r}) {1 \over r} B_{2}  +
 { \nu \over M r} ({d \over dr} +
 {2 \over r}) B_{2} +
{2\nu^{2} \over M r^{2}} B_{1}  =  + 4 M  \; M_{1}   ,
\nonumber
\end{eqnarray}
\begin{eqnarray}
  - {2 \nu \over M r }
({d \over dr} + {1 \over r}) B_{1} - {2 \nu^{2} \over M r^{2}}
B_{2}
 - {1 \over M} {d \over dr} ({d \over dr} + {2 \over r})
B_{2} -
 {2\nu \over M}{d \over dr}{1 \over r} B_{1}  = + 4 M \;
M_{2} \; , \nonumber
\end{eqnarray}
\begin{eqnarray}
2(\epsilon  + {\alpha \over r} ) B_{2}   - {2 \nu \over M r } ({d
\over dr} + {1 \over r}) B_{1}  - {2 \nu^{2} \over M r^{2}} B_{2}
+
 {1 \over M} {d \over dr} ({d \over dr} + {2 \over r}) B_{2}+
{2\nu \over M}{d \over dr}{1 \over r}  B_{1}  =  0 \; , \nonumber
\end{eqnarray}
\begin{eqnarray}
2(\epsilon+ {\alpha \over r }) B_{1}  + {1 \over M} ( {d \over dr
}+ {1 \over r } ) ^{2} B_{1}  +  { \nu \over M } ( {d \over dr } +
{1 \over r}) {1 \over r} B_{2}   -
  { \nu \over M r} ({d \over dr}
+ {2 \over r}) B_{2} - {2\nu^{2} \over M r^{2}} B_{1}   =  0  .
\nonumber
\end{eqnarray}

\noindent Two last equation  provides us with non-relativistic radial equations -- they can  be
written as
\begin{eqnarray}
 r^{2} \left [    {d^{2} \over dr^{2}}  + {2 \over r}{d \over dr}   + 2M(\epsilon  + {\alpha \over r} )
  - {2 \nu^{2} \over  r^{2}}  \right ] B_{2}  =  2   B_{2} +  4 \nu \;   B_{1}   \; ,
\nonumber
\\
 r^{2} \left [    {d ^{2} \over dr ^{2} }+ {2 \over r } {d \over dr }   +2M(\epsilon+ {\alpha \over r })
  - {2\nu^{2} \over  r^{2}}  \right ] B_{1}
 =   2\nu  \;  B_{2}     \; .
\label{6.17}
\end{eqnarray}

It is convenient to presents eqs. (\ref{6.17}) in a matrix form
\begin{eqnarray}
{1 \over 2} r^{2} \Delta \; \left | \begin{array}{c}
B_{1} \\
B_{2}
\end{array}  \right | =
\left | \begin{array}{cc}
0  &  \nu   \\
2\nu  & 1
\end{array}  \right |
\left | \begin{array}{c}
B_{1} \\
B_{2}
\end{array}  \right |.
\label{6.18}
\end{eqnarray}

\noindent The right-hand part can be brought to a diagonal form
\begin{eqnarray}
\left | \begin{array}{c}
f_{1} \\
f_{2}
\end{array}  \right | =
\left | \begin{array}{cc}
a & c  \\
d  & b
\end{array}  \right |
\left | \begin{array}{c}
B_{1} \\
B_{2}
\end{array}  \right |\; , \qquad
r^{2} \Delta \; \left | \begin{array}{c}
f_{1} \\
f_{2}
\end{array}  \right | =
\left | \begin{array}{cc}
\lambda_{1} & 0  \\
0  & \lambda_{2}
\end{array}  \right |
\left | \begin{array}{c}
f_{1} \\
f_{2}
\end{array}  \right |.
\end{eqnarray}

\noindent The problem  is to solve two systems
\begin{eqnarray}
\left | \begin{array}{cc}
a & c  \\
d  & b
\end{array}  \right |
\left | \begin{array}{cc}
0 & \nu  \\
2\nu  & 1
\end{array}  \right |
= \left | \begin{array}{cc}
\lambda_{1} & 0  \\
0  & \lambda_{2}
\end{array}  \right |
\left | \begin{array}{cc}
a & c  \\
d  & b
\end{array}  \right |,
\nonumber
\end{eqnarray}

\noindent from whence it follows
\begin{eqnarray}
\left \{ \begin{array}{l}
 \lambda_{1} \; a  - 2\nu \; c =0 \\
- \nu \; a + (\lambda_{1} -1) \; c =0 \; ,
\end{array} \right. \qquad
\lambda_{1}=  {1 + \sqrt{1 + 4j(j+1)}\over 2} = j+1 \; , \qquad
c = { \lambda_{1}   \over 2 \nu} a \; ;
\nonumber
\end{eqnarray}
\begin{eqnarray}
\left \{ \begin{array}{l}
 \lambda_{2} \; d  - 2\nu \; b =0 \\
- \nu \; d + (\lambda_{2} -1) \; b =0 \; ,
\end{array} \right. \qquad
   \lambda_{2}=  {1 - \sqrt{1 + 4j(j+1)}\over 2} = - j \;  , \qquad
 b = { \lambda_{2}  \over   2\nu }  d  \; .
\nonumber
\end{eqnarray}

\noindent The transformation matrix we need is given by
\begin{eqnarray}
\left | \begin{array}{c}
f_{1} \\
f_{2}
\end{array}  \right | =
\left | \begin{array}{cc}
a &   \lambda_{1} \; a /  2 \nu   \\
d  & \lambda_{2} \; d  /   2\nu
\end{array}  \right |
\left | \begin{array}{c}
B_{1} \\
B_{2}
\end{array}  \right |\; .
\label{6.19}
\end{eqnarray}

Thus, the system  (\ref{6.18}) is led to the diagonal form
\begin{eqnarray}
 \left [    {d^{2} \over dr^{2}}  + {2 \over r}{d \over dr}   + 2M(\epsilon  + {\alpha \over r} )
  - {2 \nu^{2} \over  r^{2}}   - {2 \lambda_{1} \over  r^{2}}  \right ] f_{1} = 0 \; ,
  \nonumber
  \\
\left [    {d^{2} \over dr^{2}}  + {2 \over r}{d \over dr}   +
2M(\epsilon  + {\alpha \over r} )
  - {2 \nu^{2} \over  r^{2}}   - {2 \lambda_{2} \over  r^{2}}  \right ] f_{2} = 0 \; .
\label{6.22}
\end{eqnarray}

\noindent
Note simple relations
\begin{eqnarray}
{2 \nu^{2} \over  r^{2}}   + {2 \lambda_{1} \over  r^{2} }=
{ j(j+1) + 2(j+1) \over r^{2}}  = {(j+1)(j+2) \over r^{2}} \; ,
\nonumber
\\
{2 \nu^{2} \over  r^{2}}   + {2 \lambda_{2} \over  r^{2}}=
{ j(j+1) - 2j  \over r^{2}}  = { (j-1) j  \over r^{2}} \; .
\nonumber
\end{eqnarray}

\noindent
Thus, we have two  problem of one the same type
(below  to $f_{1}$ and $f_{2}$ correspond $\nu= j-1$ and $\nu = j+1$ respectively)
\begin{eqnarray}
\left [    {d^{2} \over dr^{2}}  + {2 \over r}{d \over dr}   +
2M(\epsilon  + {\alpha \over r} )
  - {\nu (\nu +1) \over  r^{2}}       \right ] f = 0 \; .
\label{6.23}
\end{eqnarray}

\noindent
Changing the variable
 $x=2 \sqrt{-2\epsilon M}\,r$
\begin{eqnarray}
\left [    {d^{2} \over dx^{2}}  + {2 \over x}{d \over dx}   -{1\over 4}-{\alpha M\over x \sqrt {-2\epsilon M}}
  - {\nu (\nu+1)  \over  x^{2}}   \right ] f(x) = 0 \; ,
\label{6.24}
\end{eqnarray}

\noindent
and introducing the substitution
$
f(x)  = x^{a} e^{-bx}\; F (x)$,
eq.  (\ref{6.24}) is brought to
\begin{eqnarray}
x{d^{2}F\over dx^{2}}+(2a+2-2bx)\,{dF\over dx}+
\left[{a(a+1)-\nu(\nu+1) \over x}-2b-2ab+{\alpha M\over \sqrt{-2\epsilon M }}+(b^{2}-{1\over 4})x\right]F=0 \; .
\nonumber
\\
\label{6.26}
\end{eqnarray}

\noindent When
$b= + 1/2\, ,  \;  a = + \nu$, eq. (\ref{6.26}) is simplified
\begin{eqnarray}
x{d^{2}F\over dx^{2}}+(2\nu +2-x)\,{dF\over dx}-\left[1+ \nu -{\alpha M\over \sqrt{-2\epsilon M }}\right]F=0\; ,
\label{6.28}
\end{eqnarray}

\noindent what is  the confluent hypergeometric equation
for  $F(A,C; \; x)$  with parameters given by
$$
A=1+ \nu -{\alpha M\over \sqrt{-2\epsilon M }}\,, \qquad C=2\nu +2 \; .
$$

\noindent
The quantization condition is
$
A = -n$, which gives
\begin{eqnarray}
1+ \nu -{\alpha M\over \sqrt{-2\epsilon M }} = - n \qquad \Longrightarrow
\qquad
\epsilon = - { \alpha ^{2} M \over  2(1+\nu  + n )^{2} } = - { me^{4} \over  2 \hbar^{2}  (1+\nu  + n )^{2} } \; .
\nonumber
\\
\label{6.29}
\end{eqnarray}

\noindent remembering that  to linearly independent solutions with parity
 $P=(-1)^{j}$ correspond  $\nu = j-1$  and   $\nu = j+1$.

\section{Conclusions}

Quantum-mechanical system -- spin 1 particle in external Cou\-lomb
field is studied on the base of the matrix Duffin -- Kemmer --
Petiau formalism with the  use of the tetrad technique.
 With the help of parity
operator, the radial 10-equation system is  divided into two
subsystem of 4 and 6 equations that correspond to  parity $P=
(-1)^{j+1}$ and  $P= (-1)^{j}$ respectively. The system of 4
equation is reduced to a second order differential equation which
coincides with that arising in the case of a scalar particle in
Coulomb potential. It is shown that the 6-equation system reduces
to two different differential equations for a "main" \hspace{2mm}
function. One main equation reduces to to a confluent Heun
equation  and provides us  with energy spectrum. Another main
equation is a more complex one, and any solutions for it are not
constructed. In radial equations, transition to  non-relativistic
case is performed. In this limit, three types of linearly
independent solutions have been constructed in terms of
hypergeometric functions.

\section{Acknowledgement }

The authors are grateful to participant of the seminar of
Laboratory of theoretical physics, Institute of physics, National
academy of sciences of Belarus, for stimulating discussion.

\end{document}